\documentclass[twocolumn,floatfix,longbibliography,hyperref]{revtex4-1}
\usepackage{amsfonts,amsmath,graphicx,amssymb,bm}
\usepackage[pdftex,plainpages=false,colorlinks=true,linkcolor=blue, citecolor=blue, urlcolor=blue]{hyperref}
\pdfoutput=2

\usepackage[charter]{mathdesign}
\usepackage[x11names,svgnames]{xcolor}

\newcommand\av{\mathbf{a}}
\newcommand\rv{\mathbf{r}}
\newcommand\kv{\mathbf{k}}
\newcommand\qv{\mathbf{q}}

\newcommand\sv{\mathbf{s}}
\newcommand\Sigmav{\bm{\Sigma}}

\newcommand\Gv{\mathbf{G}}
\newcommand\Tr{\mathrm{Tr}}

\newcommand\xh{\mathbf{\hat x}}
\newcommand\yh{\mathbf{\hat y}}
\newcommand\dgf{{\phantom{\dagger}}}
\begin{document}

\title{Interplay Between $d$-wave Superconductivity and a Bond Density Wave in the one-band Hubbard model}
\author{J. P. L. Faye}
\altaffiliation{Current address: The Abdus Salam International Center for Theoretical Physics, Strada Costiera 11, 34014 Trieste, Italy}
\affiliation{D\'epartement de physique and Institut Quantique, Universit\'e de Sherbrooke, Sherbrooke, Qu\'ebec, Canada J1K 2R1}
\author{D. S\'en\'echal}
\affiliation{D\'epartement de physique and Institut Quantique, Universit\'e de Sherbrooke, Sherbrooke, Qu\'ebec, Canada J1K 2R1}
\date{\today}

\begin{abstract}

It is now well established that superconducting cuprates support a charge density wave state in the so-called underdoped region of their phase diagram. We investigate the possibility of charge order in the square-lattice Hubbard model, both alone and in coexistence with $d$-wave superconductivity. The charge order has a period four in one direction, is centered on bonds and has a $d$ form factor. We use the variational cluster approximation, an approach based on a rigorous variational principle that treats short-range correlations exactly, with two clusters of size $2\times6$ that together tile the infinite lattice and provide a non-biased unit for a period-four bond density wave (BDW).
We find that the BDW exists in a finite range of hole doping and increases in strength from $U=5$ to $U=8$. Its location and intensity depends strongly on the band dispersion. When probed simultaneously with $d$-wave superconductivity, the energy is sometimes lowered by the presence of both phases, depending on the interaction strength. Whenever they coexist, a pair-density wave (a modulation of superconducting pairing with the same period and form factor as the BDW) also exists.
\end{abstract}
\maketitle

\section{Introduction}

Charge order in underdoped superconducting cuprates has been observed by many techniques and in many compounds.
Nuclear magnetic resonance measurements on YBa$_2$Cu$_3$O$_y$ indicate the presence of a long-range, static charge order without any signature of spin order \cite{Wu2011a, Wu2013}.
In Bi$_2$Sr$_2$CaCu$_2$O$_{8+\delta}$, scanning tunnelling microscopy (STM) shows a periodic modulation in the density of states \cite{Hoffman2002a, Vershinin2004a}. 
Charge density wave correlations have also been observed in X-ray scattering \cite{Ghiringhelli2012b, Achkar2012a, Chang2012}, and the charge density wave seems to be directed along copper oxygen bonds\cite{Blackburn2013a}. 
STM measurements also indicate that the charge density wave modulation resides on Cu--O--Cu bonds \cite{Vershinin2004a, Kohsaka:2007fk, Achkar2013a}. 
The dependence of the peak  intensity as a function of magnetic field  clearly indicates the possibility of a competition between d-wave superconductivity and charge density wave order \cite{Chang2012}.
More recently, the pair-density wave (PDW) that coexists with $d$-wave superconductivity and charge order has also been observed \cite{Hamidian2016}.
 
Theoretical investigations of charge order in cuprates roughly fall into two categories: (i) those that study the effect of static charge order on observables and (ii) those that attempt at explaining the origin of charge order from a model Hamiltonian with interactions.
This work belongs to the second category. A few attempts have been made in that direction in the literature. For instance, Vojta\cite{Vojta2002a} has applied mean-field theory to the $t$-$J$ model plus extended interactions and mapped out various charge order phases that appear when $J$ is low enough, whereas $d$-wave superconductivity dominates at higher $J$. A pure exchange model (without correlated hopping) has also been studied at the mean field level by Sachdev and LaPlaca~\cite{Sachdev2013a}. Atkinson {\it et al.} have applied the generalized RPA approximation to the full three-band Hubbard model \cite{Atkinson2015} and view charge order, like the pseudogap, as a side effect of short-range antiferromagnetic correlations.
The Gutzwillwer approximation has been applied to the Hubbard model (without extended interactions) but no charge order was found with that approach\cite{Markiewicz2011}.
Charge order at half-filling in the extended Hubbard model has recently been investigated with the dynamical cluster approximation\cite{Terletska2016}, where it should be competition with antiferromagnetism. That competition has also been studied in the context of the Hubbard-Holstein model\cite{Bauer2010}, in which optical phonons would favor charge order over antiferromagnetism.

There is also a vast literature on stripe order, i.e., a coexistence of charge and spin density waves, which we will not review here.
Let us mention nonetheless the work of Corboz {\it et al.}\cite{Corboz2011} in which the nearest-neighbor $t$-$J$ model is studied using the projected-entangled pair states (PEPS) variational Ansatz, and where stripe order occurs naturally in coexistence with $d$-wave superconductivity.
This is consistent with the previous work of Capello {\it et al.}\cite{Capello2008} on the same model using variational Monte Carlo.
Finally, the pair density wave (PDW) state has been the focus of many studies \cite{Himeda2002, Raczkowski2007, Choubey2016, Freire2015, Wang2015} (for a recent review, see Ref.\cite{Fradkin2015a}).

In this work we investigate whether a particular CDW order can arise from local repulsive interactions alone, and whether it can coexist with $d$-wave superconductivity.
To this end, we apply the variational cluster approximation (VCA) \cite{Dahnken:2004} to the one-band repulsive Hubbard model.
The charge density wave studied is bond centered, has a $d$-wave form factor and will henceforth be referred to as a bond-density wave (BDW).
The VCA, and other quantum cluster methods such as Cluster Dynamical Mean Field Theory\cite{Lichtenstein:2000vn,Kotliar:2001} and the Dynamical Cluster Approximation\cite{Maier:2000a} already predict the presence of $d$-wave superconductivity in the doped one-band Hubbard model\cite{Senechal2005, Maier:2005ec,Kancharla:2008vn}.
We find that a bond density wave is indeed possible in the doped Hubbard model and that this phase is more robust when increasing the interaction strength $U$. 
Its location is also sensitive to the detailed band structure.
In addition, we find that the BDW can coexist with $d$-wave superconductivity, although both the dSC and BDW order parameters are negatively affected by their coexistence.
A pair-density wave (PDW) order also sets in when BDW and dSC orders coexist.
 
This paper is organized as follows: in Sect.~\ref{sec:model}, we describe the particular BDW studied and briefly review the VCA method; in Sect.~\ref{sec:results}, we present and discuss our numerical results. Finally, we conclude in Sec.~\ref{conclusion}.

\section{Model and method}
\label{sec:model}

Let us first establish some notation.
The one-band Hubbard model on a square lattice is defined by the following Hamiltonian:
\begin{equation}\label{eq:hubbard}
H = -t \sum_{\langle \rv,\rv' \rangle, \sigma} c^\dagger_{\rv,\sigma}c^\dgf_{\rv',\sigma} + U\sum_{\rv} n_{\rv,\uparrow}n_{\rv,\downarrow} - \mu\sum_{\rv,\sigma} n_{\rv,\sigma}
\end{equation}
where $c_{\rv,\sigma}$ destroys an electron of spin $\sigma$ at site $\rv$ on the lattice. $n_{\rv,\sigma}=c^\dagger_{\rv,\sigma}c_{\rv,\sigma}$ is the number of electrons of spin $\sigma$ at site $\rv$. 
The chemical potential $\mu$ is included in the Hamiltonian for convenience.

\begin{figure}
\centerline{\includegraphics[scale=0.8]{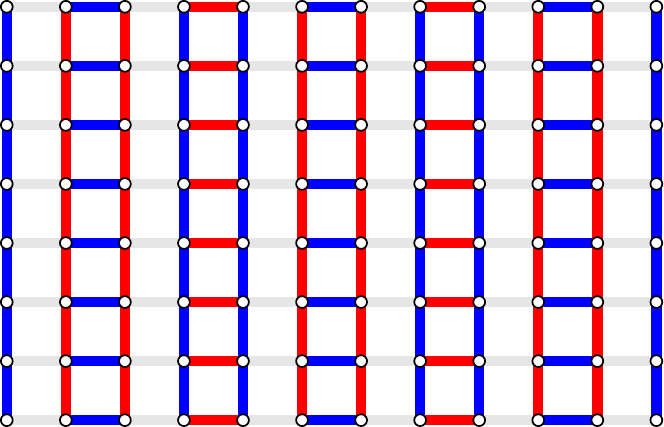}}
\caption{Bond density wave pattern studied in this work. Blue means positive, red negative and gray zero.}
\label{fig:cdw}
\end{figure}

Following  \cite{Allais2014a}, a general BDW operator of wavevector $\qv$ is defined as follows:
\begin{equation}\label{eq:BDW}
\hat\Psi_\mathrm{BDW} = 
\sum_{\rv \sigma,\av} t_{\qv,\av} c^\dagger_{\rv,\sigma} c^\dgf_{\rv+\av,\sigma} e^{i(\qv\cdot\rv+\av/2)} + \mathrm{H.c}
\end{equation}
We will probe a BDW of period four, with $d$-form factor:
$\qv=(\pi/2,0)$, with $t_{\qv,\yh} = -t_{\qv,\xh} \equiv -\lambda/\sqrt2$. The $d$-form factor is supported by STM observations~\cite{Fujita2014a}.
This BDW is illustrated on Fig.~\ref{fig:cdw}. 
This choice is motivated by its simple commensurability and its compatibility with the cluster method we will use, as explained below.

We will also probe $d$-wave superconductivity, with a pair operator defined as
\begin{equation}\label{eq:dSC}
\hat\Psi_\mathrm{dSC} = \sum_{\rv, \av=\xh,\yh} \Delta_\av \left(c_{\rv,\uparrow}c_{\rv+\av,\downarrow}-c_{\rv,\downarrow}c_{\rv+\av,\uparrow}\right) + \mathrm{H.c.}
\end{equation}
where $\Delta_\yh = -\Delta_\xh \equiv -\Delta$.
If both superconductivity and charge order are present, the pair-density wave (singlet) order parameter
\begin{equation}\label{eq:PDW}
\hat\Psi_\mathrm{PDW} = 
\sum_{\rv,\av} t_{\qv,\av} (c_{\rv,\uparrow} c_{\rv+\av,\downarrow}-c_{\rv,\downarrow} c_{\rv+\av,\uparrow}) e^{i(\qv\cdot\rv+\av/2)} + \mathrm{H.c}
\end{equation}
should also be nonzero.

\subsection{The Variational Cluster Approximation} 

In order to probe the possibility of superconductivity and bond density wave as well as their coexistence in Model \eqref{eq:hubbard}, we use the variational cluster approximation (VCA) with an exact diagonalization solver at zero temperature~\cite{Dahnken:2004}.
This method, which goes beyond mean-field theory by keeping the correlated character of the model, has been applied to many strongly correlated systems in connection with various broken symmetry phases, in particular $d$-wave superconductivity~\cite{Senechal2005,Aichhorn:2007xr}.
For a detailed review of the method, see Refs~\cite{Potthoff:2012fk,Potthoff:2014rt}.

In essence, the VCA is a variational method on the electron self-energy. It can probe various broken symmetries (or just the normal state) by exploring a space of self-energies that are the actual self-energies of Model \eqref{eq:hubbard}, but restricted on a small cluster of sites and augmented by Weiss fields that probe broken symmetries and other one-body terms.
Once the optimal self-energy in that space is found, it is added to the noninteracting Green function for the full lattice and, from there, various 
observables may be computed.

Like other quantum cluster methods, VCA starts by a tiling of the lattice into an infinite number of (usually identical) clusters. 
In VCA, one considers two systems: the original system described by the Hamiltonian $H$, defined on the infinite lattice, and the {\it reference system}, governed by the Hamiltonian $H'$, defined on the cluster only, with the same interaction part as $H$.
Typically, $H'$ will be a restriction of $H$ to the cluster (i.e., with inter-cluster hopping removed), to which various Weiss fields may be added in order to probe broken symmetries.
More generally, any one-body term can be added to $H'$.
The size of the cluster should be small enough for the electron Green function to be computed numerically.

The optimal one-body part of $H'$ is determined by a variational principle. 
More precisely, the electron self-energy $\Sigmav$ associated with $H'$ is used as a variational self-energy, in order to construct the following Potthoff self-energy functional~\cite{Potthoff2003b}:
\begin{multline}\label{eq:omega}
\Omega[\Sigmav(\xi)]=\Omega'[\Sigmav(\xi)]\\ +\Tr\ln[-(\Gv^{-1}_0 -\Sigmav(\xi))^{-1}]-\Tr\ln(-\Gv'(\xi))
\end{multline} 
The quantities $\Gv'$ and $\Gv_0$ above are respectively the physical Green function of the cluster and the non-interacting Green function of the infinite lattice. 
The symbol $\xi$ stands for a small collection of parameters that define the one-body part of $H'$. 
$\Tr$ is a functional trace, i.e., a sum over frequencies, momenta and bands, and $\Omega'$ is the grand potential of the cluster, i.e., its ground state energy, since the chemical potential $\mu$ is included in the Hamiltonian.
$\Gv'(\omega)$ and $\Omega'$ are computed numerically via the Lanczos method at zero temperature.

The Potthoff functional $\Omega[\Sigmav(\xi)]$ in Eq.~\eqref{eq:omega} is computed exactly, but on a restricted space of the self-energies $\Sigmav(\xi)$ that are the physical self-energies of the reference Hamiltonian $H'$.
We use a standard optimization method (e.g. Newton-Raphson) in the space of parameters $\xi$ to find the stationary value of $\Omega(\xi)$:
\begin{equation}\label{eq:Euler}
\frac{\partial\Omega(\xi)}{\partial\xi} = 0
\end{equation}
This represents the best possible value of the self-energy $\Sigmav$, which is used, together with the non-interacting Green function $\Gv_0$, to construct an approximate Green function $\Gv$ for the original lattice Hamiltonian $H$:
\begin{equation}\label{eq:CPT}
\Gv(\kv,\omega) = \frac1{\Gv_0^{-1}(\kv,\omega) - \Sigmav(\omega)}~~.
\end{equation}
In the above the wavevector $\kv$ is restricted to the reduced Brillouin zone associated with the superlattice defined by the cluster, and all boldface quantities are matrices of dimension $L$ (or $2L$, if superconductivity is present), $L$ being the number of sites in the cluster.
From that Green function one can compute the average of any one-body operator, in particular order parameters associated with bond density wave (BDW) or $d$-wave superconductivity (dSC).

The competition between orders can be studied by probing the two orders separately, and then together, in a coexistence scenario. If homogeneous coexistence is possible, then generally the associated value of $\Omega$, which approximates the free energy in this approach, is lower than for the pure solutions for the two phases separately.

\begin{figure}
\centerline{\includegraphics[width=\hsize]{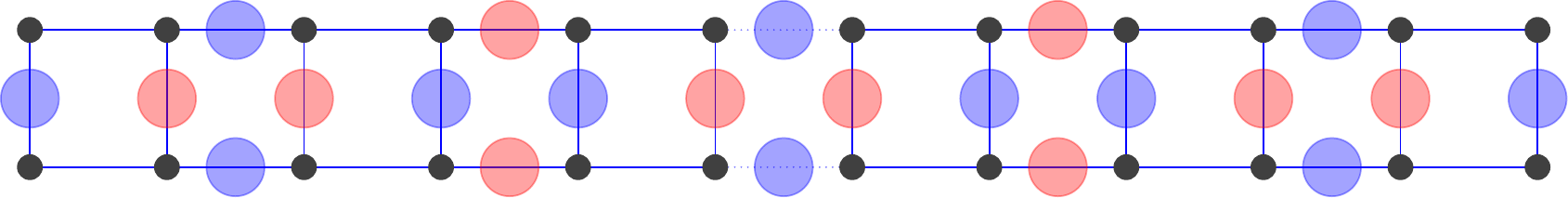}}
\caption{The two 12-site clusters used in this work (black dots) with the BDW amplitudes (blue means positive, red negative).
These two clusters together form a unit that is repeated in both $x$ and $y$ directions. Note that changing the sign of the BDW amounts to interchanging the two clusters or to flipping each cluster about the vertical axis.}
\label{fig:model-bdw}
\end{figure}

Charge density waves, or other states that break translation symmetry, present a particular challenge to cluster methods like VCA, first because the unit cell of the density wave may be larger than the largest cluster that can be practically solved numerically, and second because cluster methods breaks translation symmetry from the outset.
The first difficulty is solved by aggregating different clusters that together form a `supercluster' (or repeated unit) that tiles the lattice and that accommodates the charge density wave.
The overcome the second difficulty, one has to choose the clusters in such a way that the method is not biased towards the broken symmetry state.
In other words, if $\lambda$ and $\Delta$ are the amplitudes of the Weiss fields associated with the BDW and dSC, respectively, then the Potthoff functional $\Omega(\lambda, \Delta)$ should be an even function of $\lambda$ and $\Delta$. Thus the normal solution ($\lambda=\Delta=0$) is always an option and the broken symmetry state occurs
if a nontrivial solution exists with a lower value of $\Omega$.
In this work, we will use the two 12-site clusters shown in Fig.~\ref{fig:model-bdw};
the amplitudes of the bond charges in the particular BDW studied are indicated by colored circles.
Together, these two clusters form a supercluster that contains three unit cells of the BDW.
Changing the sign of the BDW amounts to flipping each cluster horizontally, which does not affect the
value of $\Omega(\lambda,\Delta)$.

A few words on the computation of expectation values.
The average of any one-body operator can be computed from the VCA solution in two different ways: (i) by taking its trace against the Green function or (ii) by differentiating the grand potential with respect to an external field.
Let $s_{\alpha\beta}$ be the one-body matrix defining the operator $\hat S$, such that 
\begin{equation}
\hat S = \sum_{\alpha,\beta} s_{\alpha\beta} c^\dagger_\alpha c_\beta
\end{equation}
($\alpha$ and $\beta$ are compound indices, representing site together with spin or other band indices).
In a cluster approach, a partial Fourier transform can be applied to the site indices in $s_{\alpha\beta}$
to produce a reduced expression $s_{\alpha\beta}(\kv)$, where $\kv$ belongs to the reduced Brillouin zone and the site indices are now limited to those of the repeated unit (the `supercluster'). In this language, the expectation value of $\hat S$ may be evaluated as
\begin{equation}\label{eq:trace_GF}
\langle \hat S\rangle = \int \frac{d\omega}{2\pi}\int_{\rm rBZ}\frac{d^2k}{(2\pi)^2} \mathrm{tr}\left[ \sv(\kv)\Gv(\omega,\kv) \right]
\end{equation}
where the frequency integral is taken over a contour that circles the negative real axis, targeting the occupied states only.
Alternatively, one may add an external field $s\hat S$ to the lattice Hamiltonian, compute the best estimate of the grand potential $\Omega$ for $s=0$ and $s=\epsilon$ (i.e. the optimized self-energy functional), and compute the derivative 
\begin{equation}\label{eq:external_field}
\langle \hat S\rangle = \frac{\partial\Omega}{\partial s} \approx \frac{\Omega(\epsilon)-\Omega(0)}{\epsilon}
\end{equation}
The two approaches \eqref{eq:trace_GF} and \eqref{eq:external_field} do not necessarily yield the same answer (the first one is less computationally intensive).
In the case of a local operator, like the particle number $\hat N$, which does not contain inter-cluster components, it can be shown that the two approaches will yield the same value $n = \langle \hat N\rangle$ if the corresponding Weiss field on the cluster (in this case $\mu'$, the cluster's chemical potential) is treated as a variational parameter on the same level as the others (e.g. $\lambda$ and $\Delta$).

\begin{figure}
\centerline{\includegraphics[scale=0.8]{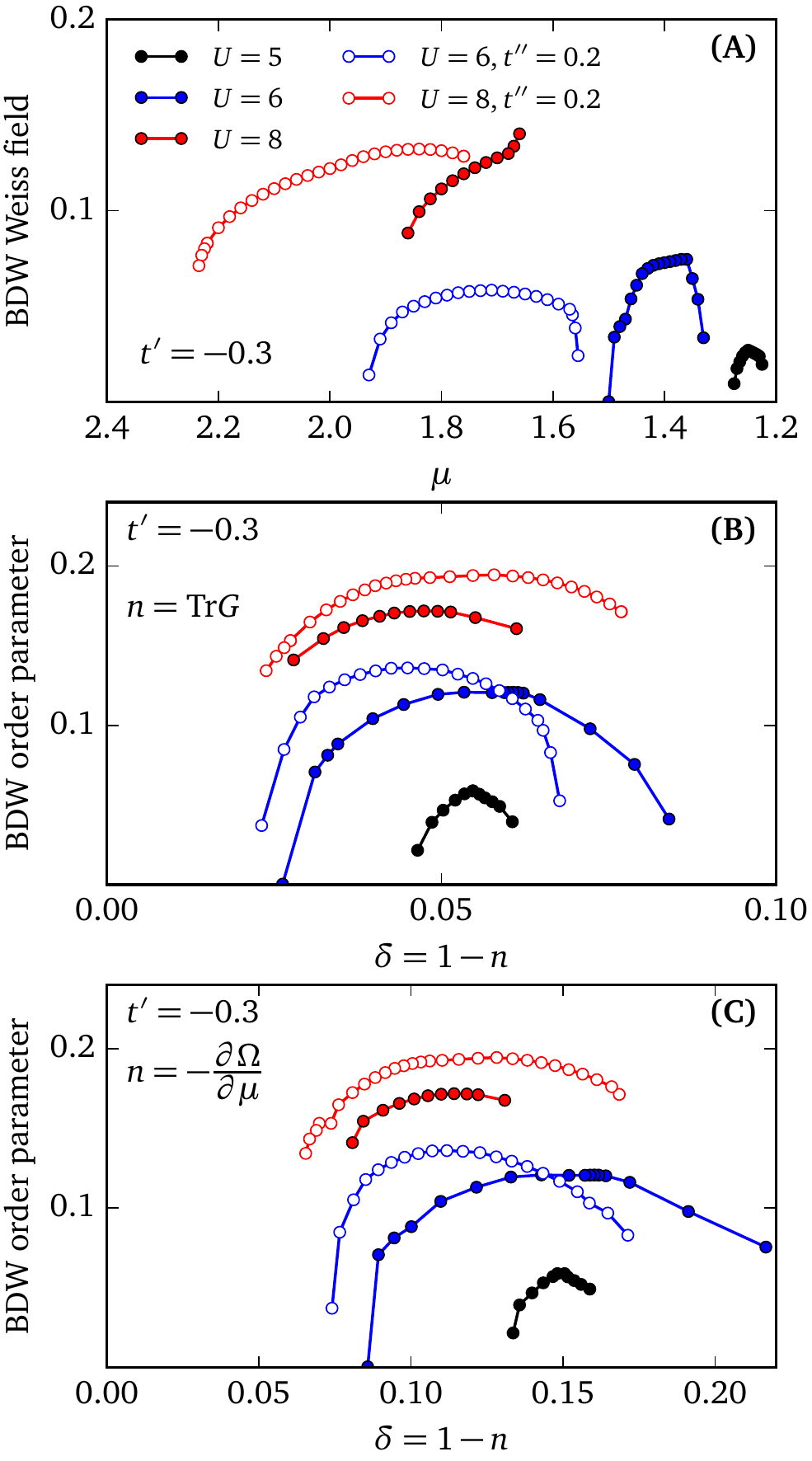}}
\caption{Top panel (A) : Optimal value of the Weiss field $\lambda$ for the pure BDW phase as a function of chemical potential $\mu$ for various values of the interaction $U$ and of the third-neighbor hopping $t''$, at $t'=-0.3$. The Weiss field goes generally to zero at the edge of the BDW phase, with the exception of $(U,t'')=(8,0.2)$.
Note the direction of the axis: larger densities are on the left.
Middle panel (B) : Corresponding BDW order parameter as a function of hole doping  $\delta = 1-n$ ($n$ being the electron density), where $n$ is computed from the Green function. Bottom panel (C) : The same, except that the electron density is computed from the derivative of the grand potential $\Omega$ with respect to $\mu$.}
\label{fig:U-cdw-tp-tpp}
\end{figure}

\section{Results and discussion}
\label{sec:results}

In this section we report the results of VCA calculations on Model \eqref{eq:hubbard} at various values of the interaction $U$ and for a few band parameters.
Since the simulations are somewhat time consuming, we could not explore the space of parameters {\it in extenso}, but our results illustrate how a bond density wave can arise at finite doping in the presence of repulsive interactions, and in coexistence with $d$-wave superconductivity also arising from the same interaction.

\subsection{Pure bond density wave}

We start by probing a pure BDW, without superconductivity.
In VCA, this amounts to solving the following Hamiltonian on the cluster system of Fig.~\ref{fig:model-bdw}:
\begin{equation}
H' = H_0' + \lambda \hat\Psi'_\mathrm{BDW}~,
\end{equation}
inserting the computed Green function into the expression \eqref{eq:omega} for the Potthoff functional, and finding the value of $\lambda$ that minimizes the functional.
In the above expression, $H_0'$ and $\hat\Psi'_\mathrm{BDW}$ are the restriction to the cluster of the kinetic energy operator and of the BDW operator \eqref{eq:BDW}, respectively.
The coefficient $\lambda$ was the only variational parameter used in optimizing the functional $\Omega$.
At this point, we have neglected the possible interaction with other phases, such as antiferromagnetism or superconductivity.

\begin{figure}
\centerline{\includegraphics[scale=0.8]{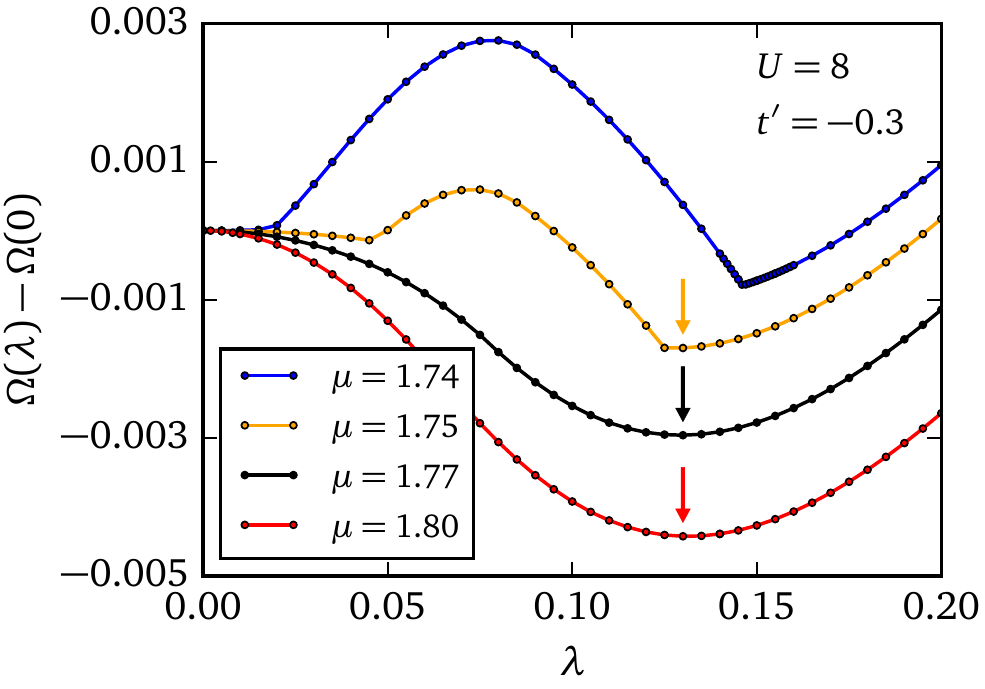}}
\caption{Self-energy functional $\Omega$ as a function of Weiss field $\lambda$ in the study of pure BDW.
Acceptable VCA solutions are indicated by arrows.
The data illustrates how the VCA solution may sometimes disappear because of the `intrusion' of segments
with different behavior and the concomitant appearance of cusp-like minima (top curve), which are not acceptable
solutions.}
\label{fig:sef}
\end{figure}

Figure~\ref{fig:U-cdw-tp-tpp}A shows the optimal value of the Weiss field $\lambda$ as a function of chemical potential $\mu$, for a few values of the local repulsion $U$, and for second-neighbor hopping $t'=-0.3$ and third-neighbor hopping $t''=0$ and $t''=0.2$. Note that the nearest-neighbor hopping $t$ is set to unity and thus defines the energy scale. 
These data constitute the `raw' solution from VCA.
It is more physically instructive to look at the order parameter $\langle\hat\Psi_\mathrm{BDW}\rangle$ as a function of electron density $n$ (or doping $\delta = 1-n$), as is done on panels (B) and (C) of the same figure.
We show the raw data on panel (A) in order to shed some light on the method itself. 

\begin{figure}
\centerline{\includegraphics[scale=0.8]{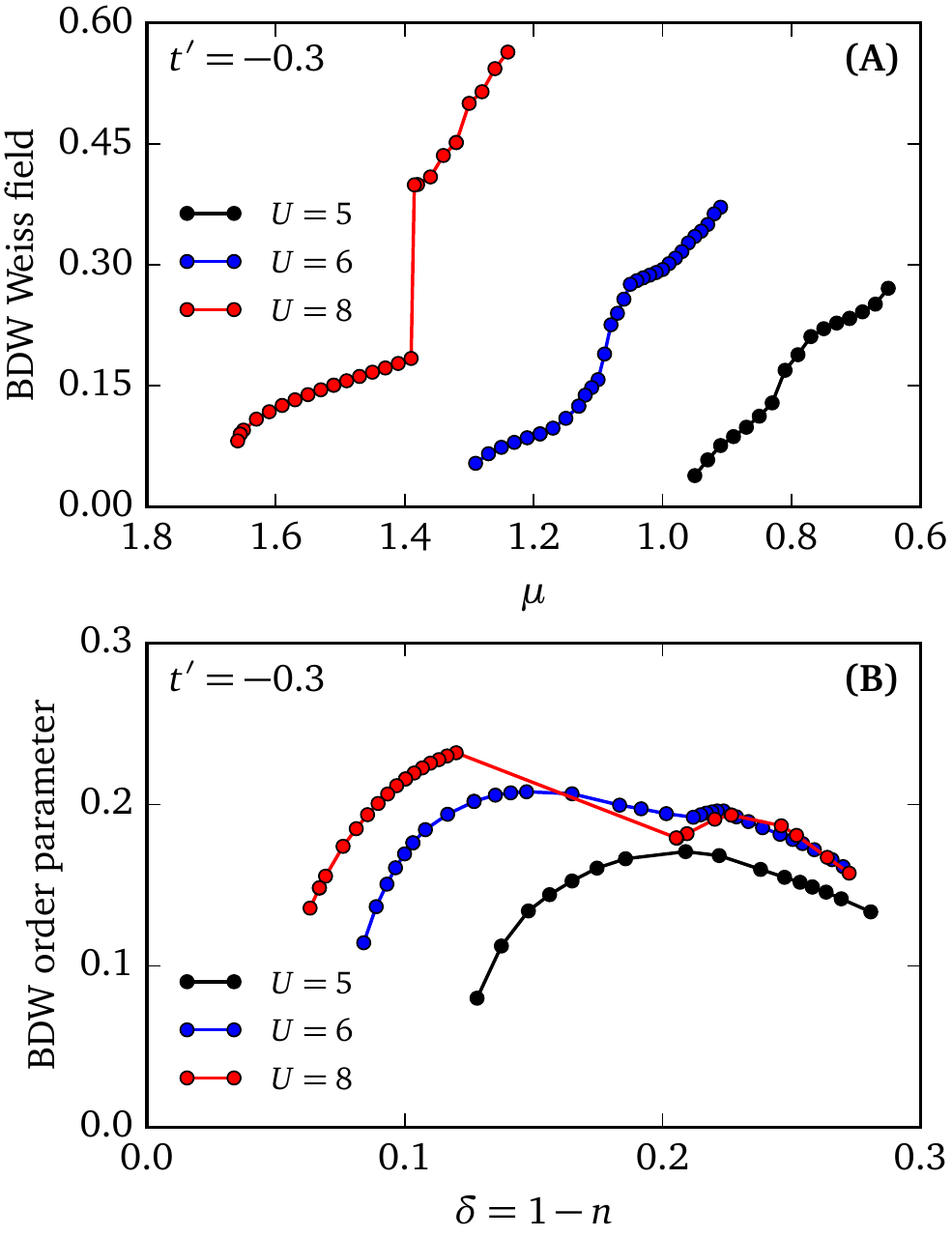}}
\caption{Top panel (A) : Optimal value of the Weiss field $\lambda$ for the pure BDW phase as a function of chemical potential $\mu$ for various values of the interaction $U$ and $t'=0$, this time by treating the cluster chemical potential $\mu'$ as a variational parameter as well. Bottom panel (A) : Corresponding BDW order parameter as a function of doping $\delta$. This time, the two methods for computing the electron density are consistent with one another.}
\label{fig:U-cdw-mu-tp}
\end{figure}

Note that the transition from the BDW phase to the normal phase can proceed in many ways: (i) the Weiss field $\lambda$ can go to zero at the phase boundary (continuous transition) or (ii) the solution can jump from a finite value of $\lambda$ to zero, either because the solution becomes metastable at that point, or because it ceases to exist (the derivative at the minimum is no longer defined, i.e., the functional $\Omega$ acquires a cusp-like behavior). The latter type occurs in the figure for $(U,t'')=(8,0.2)$. 
Fig.~\ref{fig:sef} illustrates this last point by showing plots of $\Omega(\lambda)$ for a few values of $\mu$.
We see how the smooth minimum (indicated by arrows) disappears at some value of $\mu$ as $\Omega(\lambda)$ acquires non-differentiable features.
This behavior is an occasional drawback of the method and is likely dependent on the shape and size of the cluster, on the type of Weiss field considered, etc.
In the continuous case the order parameter goes to zero smoothly, whereas it jumps discontinuously in the second case.
A more conventional first-order transition, in which two distinct {\it bona fide} solutions have the same value of $\Omega$ at some value of an external parameter, is also possible, but has not been observed here.

Panel B of Fig.~\ref{fig:U-cdw-tp-tpp} shows the BDW order parameter $\langle \hat\Psi_\mathrm{BDW}\rangle$ as a function of hole doping, both computed from the Green function (Eq.~\eqref{eq:trace_GF}), for the same data sets as those appearing on Panel A.  Panel C shows the same data, this time with the density computed from the derivative of $\Omega$ with respect to $\mu$. The values of doping obtained by these two approaches differ by roughly a factor of two.
Whatever the method of computing $n$, it appears that the BDW is increasing in strength with $U$ and is also quite sensitive on the band dispersion; in particular, no BDW order occurs at $(U,t'')=(5,0.2)$.

\begin{figure}
\centerline{\includegraphics[scale=0.8]{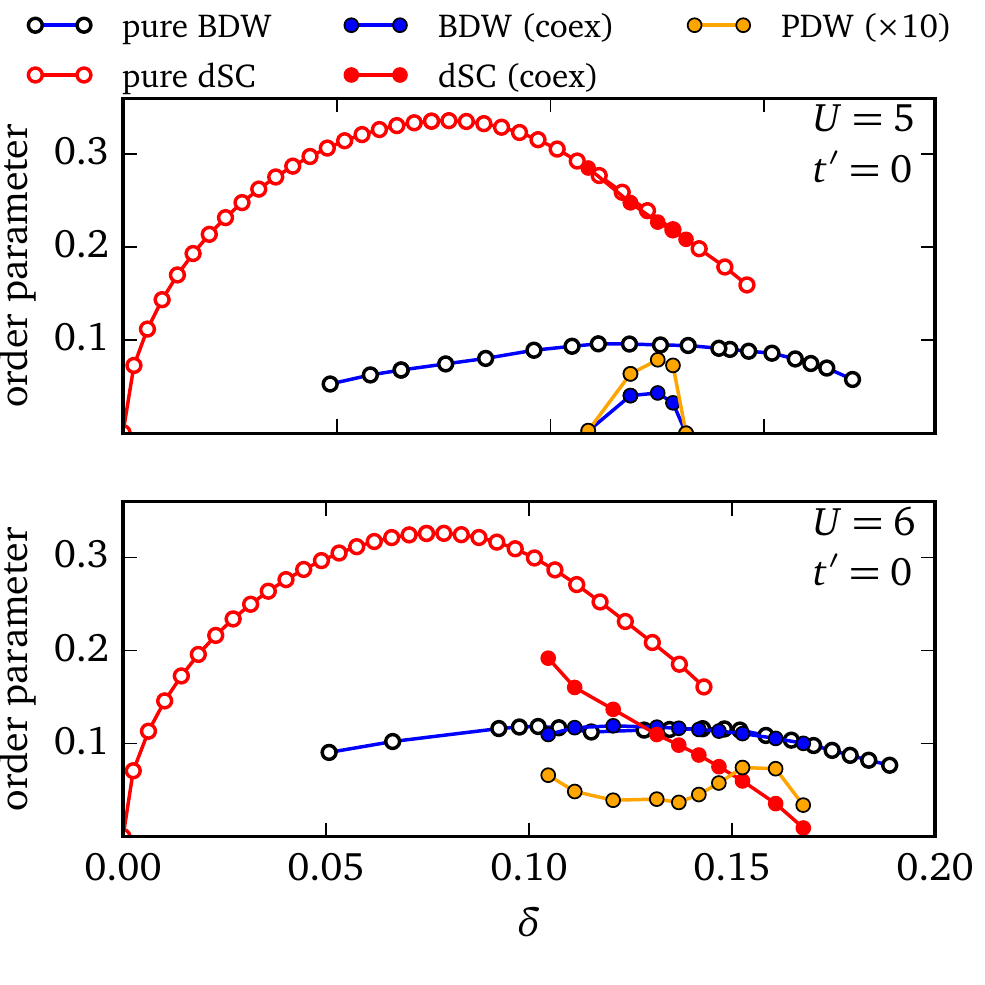}}
\caption{dSC and BDW order parameters as a function of hole doping  $\delta = 1-n$ for $U = 5$ and $U = 6$  at $t^{\prime} = t^{\prime\prime} = 0$. The pair-density wave (PDW) order parameter $\langle\hat\Psi_\mathrm{PDW}\rangle$ is also shown (note the change of scale); the latter
only exists in the coexistence phase.
}
\label{fig:U5-U6-cdw-D}
\end{figure}

\begin{figure}
\centerline{\includegraphics[scale=0.8]{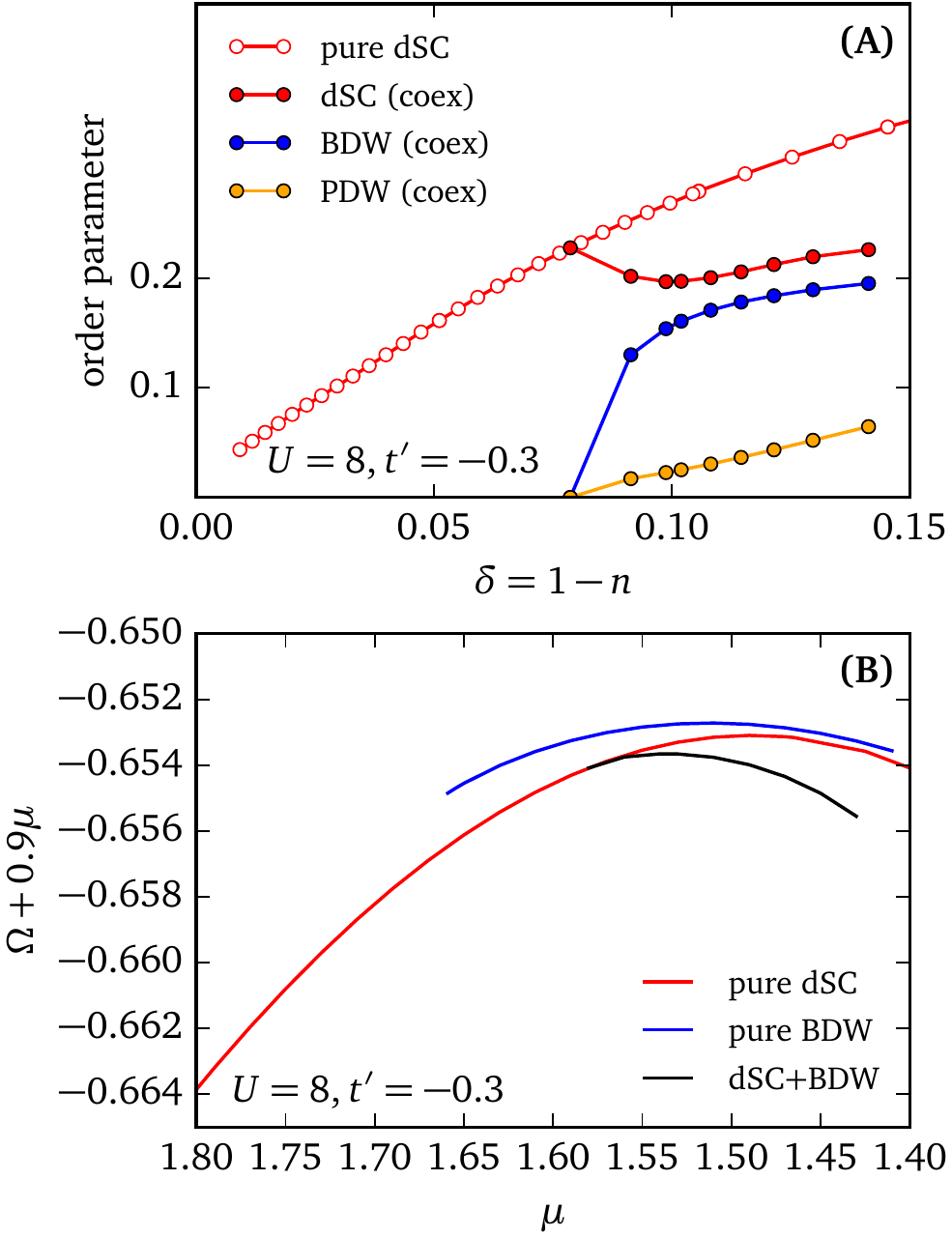}}
\caption{Top panel (A) : dSC, BDW, and PDW order parameters as a function of hole doping $\delta$ for $U=8$ and $t'=-0.3$.
The two orders were put in competition. The open red circles shows the dSC order parameter when probed alone.
In this data set the cluster chemical potential $\mu'$ was treated as a variational parameter, in addition to the Weiss fields $\Delta$ and $\lambda$. Note the dip in dSC order as the BDW order appears. Bottom panel (B): the free energy $\Omega$ of the three solutions, as a function of $\mu$ (note that the axis is reversed). The coexistence solution has the lowest energy, but exists in a narrower range of doping. The pure dSC solution always has a lower energy than the pure BDW solution shown on Fig.~\ref{fig:U-cdw-mu-tp}.}
\label{fig:U-cdw-D-mu-tp}
\end{figure}

An apparent way out of the ambiguity in the way to compute electron density is to treat the cluster chemical potential $\mu'$ as a variational parameter, on the same level as the Weiss field $\lambda$. When this is done, the data of Fig.~\ref{fig:U-cdw-mu-tp} are obtained (this was done for $t'=-0.3$ and $t''=0$ only). 
The top panel shows the Weiss field as a function of $\mu$, and the bottom panel the BDW order parameter as a function of doping.
Note that the Weiss field tends to larger values on the overdoped side (smaller $\mu$).
This leads us to think that the overdoped results are less reliable: the `canonical' phase transition in VCA has the Weiss field go
to zero at the same time as the order parameter.
A discontinuity in the solution for $U=8$, also mildly apparent at $U=6$, reinforces this point of view.

To conclude this part: Model \eqref{eq:hubbard} indeed supports a pure bond density wave as the particular one studied here, but the VCA is not too reliable as to the precise doping range where it occurs. The BDW appears quite sensitive to the dispersion relation, as expected. It appears beyond a certain threshold value of $U$ and grows with $U$ up to intermediate coupling.

\subsection{Bond density wave and $d$-wave Superconductivity}

We now proceed to probe solutions in which d-wave superconductivity (dSC) coexists with the BDW.
This is done by adding both BDW and dSC terms, as defined by Eqs~\eqref{eq:BDW} and \eqref{eq:dSC}, to the cluster Hamiltonian.
We start by studying the case with nearest-neighbor hopping only ($t'=t''=0$), for $U=5$ and $u=6$.
Fig.~\ref{fig:U5-U6-cdw-D} shows the dSC and BDW order parameters both for the pure and the coexistence solutions as a function of the hole doping $\delta$ (the `pure' solutions are obtained by setting one of the Weiss field s [$\lambda$ or $\Delta$] to zero).
The electron density was computed from the Green function.
The pure dSC solution exists in a large doping interval, with a maximum at $\delta=10\%$.
Antiferromagnetic order was not put in competition with dSC order, since we are focusing on the interplay between dSC and BDW.
The striking effect shown here is that the interaction of the two phases tends to suppress superconductivity at $U=5$, whereas it tends to suppress charge order at $U=6$.

Figure~\ref{fig:U-cdw-D-mu-tp} shows results obtained with a more realistic band dispersion ($t'=-0.3$), at $U=8$.
This time, the pure dSC solution always has a lower energy than the pure BDW solution, contrary to the solution found at $(U,t')=(6,0)$.
Nevertheless, the coexistence solution has a lower energy than the pure dSC solution from $\delta\sim9\%$ on.
We do not believe the solutions beyond $\delta=15\%$ (not shown) to be reliable, as they display the same type of discontinuity as shown on Fig.~\ref{fig:U-cdw-mu-tp}.
Note that the dSC solution reconnects to the pure-dSC solution when the BDW falls to zero, at $\delta\sim 8\%$. 

The VCA approach used in this work has strengths and weaknesses. The main strength is that the problem can be treated at all: In the context of the intermediate couping, repulsive Hubbard model, VCA and other quantum cluster methods (CDMFT, DCA) are among the few options capable of revealing $d$-wave superconductivity; the effective pairing interaction is dynamically generated within the clusters used. In addition, treating a period-four BDW can be done elegantly with $2\times6$ clusters, which are too large to be treated with an exact diagonalization solver (hence at zero temperature) in any other approach than VCA. Seeing both BDW and dSC emerging from the same repulsive interaction, in coexistence, is particularly satisfying.

On the flip side, VCA, contrary to DMFT-like approaches, does not let particles in and out of the cluster (there are no bath degrees of freedom) and this leads to a less reliable estimate of the electron density. The approach is prone to (likely spurious) first-order phase transitions that occur when doping is pushed too far. In addition, the approach is numerically more delicate than DMFT and convergence may be more difficult. Finally, in practice, it can only probe orders that are defined from the outset (i.e. it is somewhat restricted).

\bigskip
\section{Conclusion}\label{conclusion}

Figure~\ref{fig:U-cdw-D-mu-tp} is the most significant result of this work. It shows how a particular bond-density wave and d-wave superconductivity both emerge dynamically from the one-band Hubbard model and how they are intertwined in a finite range of hole doping, for a realistic dispersion relation.

In this work only local interactions are responsible for the establishment of both charge order and $d$-wave superconductivity.
It is reasonable to expect that extended interactions will reinforce the tendency towards charge order; in the context of quantum cluster methods, extended interactions need to be treated partly in the Hartree approximation (the so-called dynamical Hartree approximation \cite{Faye:2015kx}).
Work in that direction would be the natural next step.

\begin{acknowledgments}
We thank A.-M. Tremblay for fruitful discussions.
Computing resources were provided by Compute Canada and Calcul Qu\'ebec.
This research is supported by NSERC grant no RGPIN-2015-05598 (Canada) and by FRQNT (Qu\'ebec).
\end{acknowledgments}

%

\end{document}